\documentclass[article]{aa} % for a article version
\usepackage{graphicx}
\usepackage{txfonts}

\newcommand\simlt{\lower.5ex\hbox{$\; \buildrel < \over \sim \;$}}
\newcommand\simgt{\lower.5ex\hbox{$\; \buildrel > \over \sim \;$}}
\newcommand\kms{km\,s$^{-1}$}

\newcommand\halpha{H$\alpha$\,\,}

\newcommand\be{\begin{equation}}
\newcommand\ee{\end{equation}}

\newcommand{\Bbeta}{\mbox{\boldmath$\beta$}}

\newcommand\apj{ApJ}
\newcommand\aap{A\&A}
\newcommand\nat{Nature}
\newcommand\mnras{MNRAS}

\begin{document}

\title{Relativistic MHD simulations of pulsar bow-shock nebulae}

\author{
N. Bucciantini \inst{1}, E.Amato \inst{1}, L.Del Zanna \inst{2}
}
\offprints{N.Bucciantini\\
 e-mail: niccolo@arcetri.astro.it}

\institute{I.N.A.F. Osservatorio Astrofisico di Arcetri,
 Largo E.~Fermi 5, I-50125 Firenze, Italy
\and
Dip. di Astronomia e Scienza dello Spazio,
  Universit\`a di Firenze, Largo E.~Fermi 2, I-50125 Firenze, Italy}

\date{Received  19 October 2004 / Accepted  13 December 2004}
\abstract{Pulsar bow-shock nebulae are a class of pulsar wind nebulae (PWNe) that form when the pulsar wind is confined by the ram pressure of the ambient medium, and are usually associated with old pulsars, that have already emerged from the progenitor Supernova Remnant (SNR). Until a few years ago these nebulae were mainly observed as \halpha sources; recently, also non-thermal emission has been detected. This is the signature of accelerated particles gyrating in a magnetic field. In the same way as \halpha radiation is a tool for studying the layer of shocked Interstellar Medium (ISM), the non-thermal radiation might be used to infer the properties of the shocked pulsar wind. However theoretical and numerical models have been presented so far only in the hydrodynamical (HD) regime, while in order to properly model the internal flow structure and the emission properties of these nebulae a magnetohydrodynamical (MHD) treatment is required. We present here relativistic MHD (RMHD) axisymmetric simulations of pulsar wind bow-shock nebulae. The structure and fluid dynamics of such objects is investigated for various values of the pulsar wind magnetization. Simulated synchrotron maps are computed and comparison of the emission pattern with observations is discussed. 
}
\titlerunning{RMHD simulations of pulsar bow-shock nebulae}
\maketitle

%%%%%%%%%%%%%%%%%%%%%%%%%%%%%%%%%%%%%%%%%%%%%%%%%%%%%%%%%%%%%% 1
\section{Introduction}

Since the publication of the earliest sample of pulsars (Gunn \& Ostriker \cite{gun70}), it has become apparent that many of them are characterized by a high spatial velocity. With the increase of the number of known pulsars, and the improvement in the determination of their distance and position, also the largest inferred velocity for this class of objects has increased and is presently of order $1000$ \kms (Cordes \& Chernoff \cite{cordes98}). The velocity dispersion is also noticeable, since it seems to exceed that of any other galactic stellar population. An analysis by Hansen \& Phinney (\cite{hansen97}) estimates a dispersion of order $200$ \kms, while more recent studies (Arzoumanian et al. \cite{arzoumanian02}) show strong evidence for a double peaked distribution, with characteristic peak velocities of $\simeq 90$ \kms and $\simeq 500 $ \kms. In spite of the large uncertainties that still affect it, the latter result leads to estimate that up to 10\% of the pulsars younger than 20 kyr are located outside the SNR left by their progenitor.

Pulsars are known to be source of ultrarelativistic magnetized winds (see e.g. Michel \& Li \cite{michel99}), mainly made of pairs and a toroidal magnetic field. Given the high Lorentz factor with which the wind is expected to expand, in the range $10^4-10^7$ (Kennel \& Coroniti \cite{kennel84}), its interaction with the ambient medium creates a region of hot shocked plasma, that may be revealed as a source of synchrotron and Inverse Compton radiation. Emission of this nature is observed in plerions, where a young pulsar is still inside the shell of the progenitor SNR: the best known examples are the Crab and Vela nebulae.

In the case of a pulsar moving through the ISM, the interaction of its wind with the ambient medium gives rise to a cometary like nebula (Bucciantini \& Bandiera \cite{bucciantini01}), similar to what is expected for the solar wind heliosphere (Zank \cite{zank99}). Typical velocities of pulsars are much higher than the sound speed in the ISM, which ranges from $\sim 10$ \kms, for a $10^4$ K partially ionized medium, to $\sim 50$ \kms, for a warm $10^5$ K ionized one. Given the typical values of proper motion, a pulsar drives a bow-shock in the ISM, that might be revealed as a \halpha nebula, with emission mainly due to charge-exchange and collisional excitation of neutral hydrogen in the bow shock itself. 

Only five pulsar wind bow-shock nebulae have been observed in \halpha so far: PSR 1957+20 (Kulkarni \& Hester \cite{kulkarni88}), PSR 2224+65, the Guitar Nebula (Cordes et al. \cite{cordes93}), PSR J0437+4715 (Bell et al. \cite{bell95}), PSR 0740-28 (Jones et al. \cite{jones01}), and PSR 2124-3358 (Gaensler et al. \cite{gaensler02b}). In addition, it is not clear yet whether the cometary nebula associated with the isolated neutron star RX J185635-3754 (van Kerkwijk \& Kulkarni \cite{kerkwijk01}) results from a bow-shock. Finally, the recently detected small \halpha nebula coincident with the position of AX J0851.9-4717.4, located at the center of SNR G266.2-1.2 (Pellizzoni et al. \cite{pellizzoni03}), has also been interpreted as the signature of a new bow-shock nebula.

In analogy with plerions, one might in principle expect bow-shock PWNe to shine also at radio and X-ray frequencies through synchrotron radiation. X-ray emitting electrons are a powerful probe of the fluid dynamics because their short lifetime allows one to trace the flow structure inside the nebula and to constrain the pulsar properties. However \halpha bow-shocks are usually associated with old pulsars, whose luminosity has often fallen below the threshold for which the associated PWN could be detected in X-rays. It is thus not surprising that significant advances in detecting non-thermal emission from such nebulae have been made after the launch of the {\it Chandra} X-ray Observatory.

Some interesting examples of elongated non-thermal nebulae associated with pulsars still inside the SNR or just emerging have been recently found. IC443 (Bocchino \& Bykov \cite{bocchino01}) shows an elongated shape and a map of spectral index suggests it results from a pulsar moving in its SNR. There has been no pulsar detection so far, even if there is evidence for an excess of thermal X-ray emission that can be interpreted as the signature of a neutron star. Another example of a more distorted nebula is W44 (Petre et al. \cite{petre02}), where the pulsar is at the tip of a narrow protuberance connecting it to the main body of the PWN. In CTB80 the pulsar PSR B1951+32 is located just inside the outer edge of the SNR (Fesen et al. \cite{fesen88}). There is evidence, from synchrotron emission, for a PWN inside the remnant, but the position of the pulsar lays just on the edge of a narrow emission finger that protrudes from the main body of the PWN. This was interpreted assuming a higher speed and the interaction with the denser SNR shell material. 

A possible transition case between pulsars inside and outside their SNR is that of G5.4-1.2, although recent measurements of the pulsar proper motion have questioned the PSR-SNR association. Here the pulsar, that appears to have just emerged from the SNR, is at the tip of a flat spectrum radio protuberance just outside the SNR shell (Frail \& Kulkarni \cite{frail91}). Coincident with such radio finger there is a X-ray tail that is likely to be synchrotron radiation from the shocked pulsar wind (Kaspi et al. \cite{kaspi01}). The X-ray nebula does not seem less extended than the radio tail, thus suggesting that the fluid speed must be high enough so that the flow time is shorter than the high energy electrons lifetime.

The first detection of X-ray emission from a bow-shock nebula interacting with the ISM was associated with the Guitar Nebula (Romani et al. \cite{romani97} based on {\it ROSAT} data). However recent {\it Chandra} observations have shown the presence of a faint X-ray filament in the field (Chatterjee \cite{chatterjee04}) which is clearly not associated with the nebula, and which might be at the origin of that first detection. Recently, X-ray emission has been detected in the vicinity of PSRB1957+20 (Stappers et al. \cite{stappers03}), a pulsar already known to be associated with a \halpha bow-shock nebula. There is a bright extended X-ray source coincident with the position of the pulsar itself and this has been interpreted as the signature of the wind termination shock (TS). A X-ray tail aligned to the bow-shock axis is seen to extend from this nebula in the direction opposite to the pulsar proper motion. However, perhaps the most impressive non thermal bow-shock nebula, due to the high quality of both radio and X-ray data, is G359.23--0.82, the Mouse Nebula (Gaensler et al. \cite{gaensler04}). Again the X-ray tail is very extended compared to the assumed size of the bow-shock. The X-ray emission peaks close to the pulsar, where the TS is expected to be. Radio imaging of the tail seems also to suggest a structured morphology with a fainter and larger envelope.
  
In all these cases it seems that the X-ray tail has an almost cylindrical shape and a smaller transverse section than what expected based on analytic models of bow-shocks. Moreover observations seem to suggest that the flow velocity in the tail must be much higher than the proper motion of the pulsar, and possibly close enough to the speed of light to give appreciable beaming.

A different case is that of Geminga, where two faint tails of emission, extending in a direction opposite to the pulsar proper motion, have been revealed by {\it Chandra} (Caraveo et al. \cite{caraveo04}). This emission might be connected with a possible bow-shock nebula, despite the fact that Geminga is not a typical radio pulsar, and that the two tail morphology is different from what is observed in other bow-shock nebulae both outside and inside the SNR.

Apart from PSR B1957+20, in no other case non-thermal and \halpha emission have been detected together. This dichotomy might be explained considering the efficiency for producing these two radiation components and how it evolves with the age of the pulsar. Young pulsars have a higher $\dot{E}$, favoring the detection of non-thermal emission, but they have a higher surface temperature (Tsuruta \cite{tsuruta86}), and the photo-ionizing flux is usually high enough to completely pre-ionize the ISM in front of the bow-shock. Older pulsars are usually much colder, so that neutrals can survive (Bucciantini \& Bandiera \cite{bucciantini01}), but they are characterized by a lower value of $\dot{E}$ (we are not considering here millisecond pulsars that belong to binary systems) so that even if a synchrotron nebula is present the emission easily falls below the threshold of detectability. 

Although it is known that pulsar winds are magnetized, and that the ratio between the Poynting and kinetic energy flux in the wind can in principle be as high as 1 (see i.e. Arons \cite{arons02} for a discussion on pulsar wind magnetization and the $\sigma$ problem), models for these nebulae have been presented only in the classical HD regime (Bucciantini \cite{bucciantini02}, van der Swaluw et al. \cite{swaluw03}). This can be considered a good approximation for the outer layer of shocked ISM, where the \halpha emission comes from: here one does not expect the magnetic field to play a dominant role given the typical low magnetization of the ISM. However the neglect of magnetic fields, as well as the non relativistic assumption, in principle, prevent one to use those models to infer the properties of the relativistic shocked wind, and to obtain more realistic results to compare with observations, as far as non-thermal emission is concerned. 

The general structure of the fluid dynamics in pulsar bow-shock nebulae is a 3D problem. There are two preferential directions: the pulsar velocity direction, which defines the ``symmetry'' axis of the cometary nebula; and the spin axis of the pulsar, which defines the direction of the wound-up magnetic field. Additional 3D effects may arise also as a consequence of anisotropic pulsar wind energy flux (Bogovalov \cite{bogovalov99}, Komissarov \& Lyubarsky \cite{komissarov03}, Del Zanna et al. \cite{delzanna04}). The complete treatment of the system in 3D is at present computationally prohibitive. The difficulty is intrinsically connected with the presence of two very different time scales. The time scale for the relativistic wind component, which defines the time step, is much smaller than that of the non relativistic ambient medium, which defines the global evolutionary time scale of the nebula. A simplification is possible if one assumes that the spin axis is aligned with the pulsar velocity direction. In such case the problem is reduced to an axisymmetric geometry which can be easily modeled in 2D.

We want to stress that such an approximation, in addition to being supported by observations of the Vela and Crab pulsars (Helfand et al. \cite{helfand01}), can be also justified on the basis of the models that until now have been proposed for the origin of the pulsar proper motion. These models can be essentially divided in three categories, depending on the origin of the kick: supernova explosion asymmetries (Lai \& Goldreich \cite{lai00}), disruption of a binary system (Tauris \& Takens \cite{tauris98}), MHD rocket effect (Harrison \& Tademaru  \cite{harrison75}). In general, if the source of the kick lasts longer than the core collapse (1 ms) one expects the spin axis to end up parallel to the direction of motion of the pulsar itself (Spruit \& Phinney \cite{spruit98}). 

In the following we present the results of a series of 2D relativistic MHD simulations of these systems, focusing on the effects of the degree of magnetization of the pulsar wind. In Sect.~2 we describe the setup and the initial conditions of our numerical models, as well as discussing the limitations intrinsic to our approach. In sect.~3 we discuss the changes in the flow structure and in the behavior of the various fluid quantities for increasing values of the magnetization parameter. In Sect.~4 we compute simulated emission maps resulting from the flow structure we find. We discuss the general spatial properties of the synchrotron emission in these systems and how the extent of the nebula, is related to the wind magnetization. A comparison with observations in the case of the Mouse Nebula is presented. Finally, in Sect.~5 we summarize our conclusions.

%%%%%%%%%%%%%%%%%%%%%%%%%%%%%%%%%%%%%%%%%%%%%%%%%%%%%%%%%%%%%% 2
\section{Simulation setup and initial conditions}

Simulations are performed using the shock-capturing  central scheme for relativistic MHD developed by Del Zanna \& Bucciantini (\cite{delzanna02}) and Del Zanna et al. (\cite{delzanna03}), to which the reader is referred for the numerical algorithms employed. If one assumes the pulsar spin axis to be parallel to the pulsar direction of motion, the flow dynamics can be reduced to a 2D axisymmetric configuration in the frame comoving with the pulsar. In this case one can use a cylindrical grid ($R,z,\phi$). A further simplification of the problem can be achieved if one assumes $V_{\phi}\equiv B_{z}\equiv B_{R}\equiv 0$. In this case the number of fluid variables evolved by the code is reduced from 8 to 5, and the fluid equations are simplified (see also Del Zanna et al. \cite{delzanna04}). Symmetric reflection conditions are imposed on the axis, while continuous zeroth-order extrapolation is imposed on the other boundaries of the computational box. We also assume both for the pulsar wind and ambient medium the same adiabatic index 4/3 (see Bucciantini \cite{bucciantini02} and Bucciantini et al. \cite{bucciantini04} for multifluid simulations with two different adiabatic coefficients and the related problems).

A spherically symmetric relativistic wind  is injected from location $(R=0,z=0)$ representing the pulsar position, with the following characteristics. The wind Lorentz factor is $\gamma=10$; the wind is cold, with $\rho c^2/p= 100$. In order to evaluate how the flow structure changes with the wind magnetization, three values of the magnetization parameter $\sigma=r^2B(r)^2c/\dot{E}$ are adopted, namely $\sigma=0.2,0.02,0.002$. The toroidal field, wrapped around the symmetry axis, can be assumed to have the same polarization on both hemispheres since the MHD equations, in our simplified 2D axisymmetric setting, only depend on $B^2$. The ambient medium is modeled as a uniform unmagnetized wind entering our computational box with a speed $V=0.03 c$  and Mach number $M=30$. Even if the wind speed is about one order of magnitude higher than the proper motion of the fastest known pulsar, the outer medium still evolves as a non-relativistic fluid. This choice is only made to allow a faster evolution of the nebula without changing significantly the flow dynamics with respect to a case where a typical value of the pulsar speed is used.

As it is known both from analytic theory (Wilkin \cite{wilkin}) and from classical HD simulations (Bucciantini \cite{bucciantini02}, van der Swaluw \cite{swaluw03}), bow-shock nebulae are characterized by a typical length scale, the so called stagnation point distance, $d_o$. This is the distance from the pulsar at which the wind momentum flux is balanced by the ISM ram pressure: 
\be
d_o=\sqrt{\dot{E}/(4\pi c\rho_oV^2)}
\label{eq:stag}
\ee
with $\rho_o$ the density of the ISM. This means that the dimensions of the nebula are a function of the total energy flux from the pulsar, and the ram pressure of the ambient medium only. We have chosen a grid with uniform resolution ($\Delta R =\Delta z$) and a number of cells such that $d_o/\Delta z =80$. In a HD case this corresponds to having about 100 cells along the axis between the pulsar and the contact discontinuity (CD) position (Bucciantini \cite{bucciantini02}), whereas this number of cells increases with $\sigma$ as we will discuss in the next section.

The axisymmetric assumption, which provides an essential simplification of the problem, forces the magnetic field to be exclusively toroidal and wrapped around the symmetry axis. The reduction of the degrees of freedom of the system leads to unphysical over-stability, that has important implications on the dynamics especially in the head of the nebula. Here, due to hoop stresses, magnetic field loops tend to pile up. For low magnetizations this has the effect of unrealistically deforming the very head of the system, close to the axis, into a cone-like shape.  With increasing values of $\sigma$, no ``quasi-steady'' solution can be found, and the CD is pushed out of the computational box. We considered increasingly large computational domains, up to the point of placing the left edge of the computational box as far as $25d_o$ from the pulsar in the case $\sigma=0.02$. This fact did not solve the problem but allowed us to measure the speed at which the CD moves out of the box at several different times. We find that this speed (which is a fraction of the speed of light) does not decrease in time, contrary to what one would expect if a steady state were to be reached at some point. We interpret this fact as a result of magnetic collimation: when the latter becomes important the post-shock flow in the vicinities of the axis is not allowed to diverge laterally contrary to the requirements for the existence of a steady state solution.

In a real multidimensional case, kink instabilities will eventually arise (Begelman \cite{begelman98}), destroying the symmetry of the problem, partly dissipating magnetic field. In our simplified geometry, where this kind of instability is not allowed to grow, we decided to artificially dissipate magnetic fields. Our choice is justified by the expectation that kink instability will eventually arise in a strong magnetic filed, however the dissipation we introduced has no pretence of reproducing the realistic development of kink instability. In the case with $\sigma=0.002$ it is possible to find a quasi-steady state solution without imposing any dissipation. Whereas for the case with $\sigma=0.2$ we had to modify the wind structure by introducing an unmagnetized conical region: within an angle of 20 deg from the axis in the forward direction the wind is unmagnetized, though its energy flux is the same as everywhere else in the solid angle; moreover the magnetic field strength is set to zero also downstream of the TS, mimicking magnetic dissipation. To allow direct comparison among the various cases we have investigated, we decided to adopt the modified wind structure for all values of $\sigma$.

%%%%%%%%%%%%%%%%%%%%%%%%%%%%%%%%%%%%%%%%%%%%%%%%%%%%%%%%%%%%%% 3
\section{Flow structure}

The geometry and internal flow structure of the bow-shock nebula are shown in Figs.~\ref{fig:st1}-\ref{fig:st3} for the three values of the wind magnetization we adopted, namely $\sigma=0.002,0.02,0.2$. 

\subsection{Outer layer}

First of all let us discuss the shape of the external layer, which is where \halpha emission is believed to originate. We notice that this does not change significantly with the wind magnetization. Major deviations are present only in the head where hoop stresses and magnetic pinching are stronger. Without magnetic dissipation, the piling up of magnetic loops drives the front of the bow-shock to move further away from the pulsar along the axis, causing the very head to assume a rather conical shape. When artificial dissipation is included, the global shape (especially in the tail) does not show sensible variations as the wind magnetization increases. However the front shock distance from the pulsar (along the axis) can change significantly. 

In the HD case, it can be proved that, if both the ambient medium and the stellar wind have the same adiabatic coefficient, the stagnation point distance Eq.~\ref{eq:stag} exactly defines the distance of the wind TS from the pulsar along the axis in the forward direction (FTS). However, even if the adiabatic coefficients of the two media are different, $d_{\rm FTS} \simeq d_o$ still holds as an approximate relation. The positions of the CD and of the front bow-shock depend on the fluid properties (i.e. compressibility) of the wind and ambient medium respectively. An ordered magnetic field is less compressible than a gas and its pressure, rather than increasing above equipartition, pushes the front bow-shock further away from the pulsar as $\sigma$ increases.
 
It is interesting to notice that there is a ``critical'' polar angle for hoop stresses: streamlines originating closer to the axis are collimated toward it while those originating at larger polar angles are advected in the tail. We found that such critical angle increases with magnetization and its value is $5^\circ$, $20^\circ$ and  $80^\circ$ for $\sigma=0.002,0.02,0.2$ respectively.

The increase of the distance of the front shock from the pulsar has consequences that go beyond changing the length-scale of the system. It affects both the microphysical properties of the outer layer and its \halpha emission, and the inferred conditions of the ambient medium. While the pulsar spin down energy and its proper motion can be measured directly, the front shock distance is used to infer the ISM density. But increasing the magnetization of the wind from $\sigma=0$ to $\sigma=0.2$, we find that such distance increases by a factor $\sim 3$: consequently the inferred outer density might be underestimated by about one order of magnitude, if it is computed based on the HD model. The same kind of mistake could affect the estimate of the density of the neutral component based on \halpha emission, given that the surface density of the external layer, which determines the fraction of interacting neutrals, scales as $\rho_{o}d_o$  (Bucciantini \cite{bucciantini02b}).

\subsection{Pulsar wind cavity}

Let us focus now on the internal portion of the nebula occupied by the magnetized relativistically hot plasma coming from the pulsar wind. It has been suggested that the shape of the TS can be observed in the X-rays. For instance, Stappers et al. \cite{stappers03} interpret the luminous knot around PSRB1957+20 as the TS itself, while in the case of the Mouse Nebula Gaensler et al. (\cite{gaensler04}) suggest the identification of the TS with the X-ray tongue. This latter work shows that the elongation of the X-ray emitting region agrees with the shape of the TS one could expect based on a classical HD model. In previous numerical studies, Bucciantini (\cite{bucciantini02}) and van der Swaluw et al. (\cite{swaluw03}) assumed that asymptotically the pressure in the tail would reach the value of the ISM pressure, and found the elongation of the TS (ratio between the distance of the forward and backward parts of the TS) to be proportional to the Mach number $M$. In the simulation by Gaensler et al. (\cite{gaensler04}), instead, the backward TS (BTS) never moves further from the pulsar than 5-6 times the FTS distance, for whatever value of the Mach number. Our simulations confirm that the average pressure in the tail seems to saturate at a value $P_{\rm tail}\sim 0.02\rho_oV^2$. The detailed pressure profiles in the $z$-direction show a shallow gradient: this leaves open the possibility that pressure equilibrium with the ISM is reached, at large distances. However the elongation of the TS seems to depend strongly on the magnetization of the nebula. With increasing magnetization the BTS moves toward the pulsar and the back of the free flowing wind cavity shrinks. Concerning then the FTS, in our simulations this keeps approximately the same distance from the pulsar for all values of $\sigma$. However the effect of the artificial dissipation we introduced is certainly important for the fluid dynamics in the head, hence no definite conclusion can be drawn on the position of the FTS, although we expect it to be less affected by the magnetization, than the BTS, since, as stressed before, its position is determined mainly by ram pressure equilibrium.

Also the shape of the BTS changes with magnetization. In HD cases it is concave toward the pulsar and the shock is perpendicular to the wind speed (Bucciantini \cite{bucciantini02}). As $\sigma$ increases the shock becomes convex. This is a consequence of the radial profile of the total pressure in the nebula. When the azimuthal magnetic field is present the internal pressure is non-uniform in the $R$ direction, tending to be higher toward the axis (e.g. Begelman \& Li \cite{begelman92}): this causes the BTS on the axis to be closer to the pulsar than at larger cylindrical radii. As the BTS becomes more and more convex it also becomes highly oblique and the parallel (to the shock) component of the velocity becomes a non negligible fraction of $c$.

\subsection{Inner flow structure and tail}

\begin{figure}[h!!!!!!!!!!!!!]
\resizebox{\hsize}{!}{\includegraphics{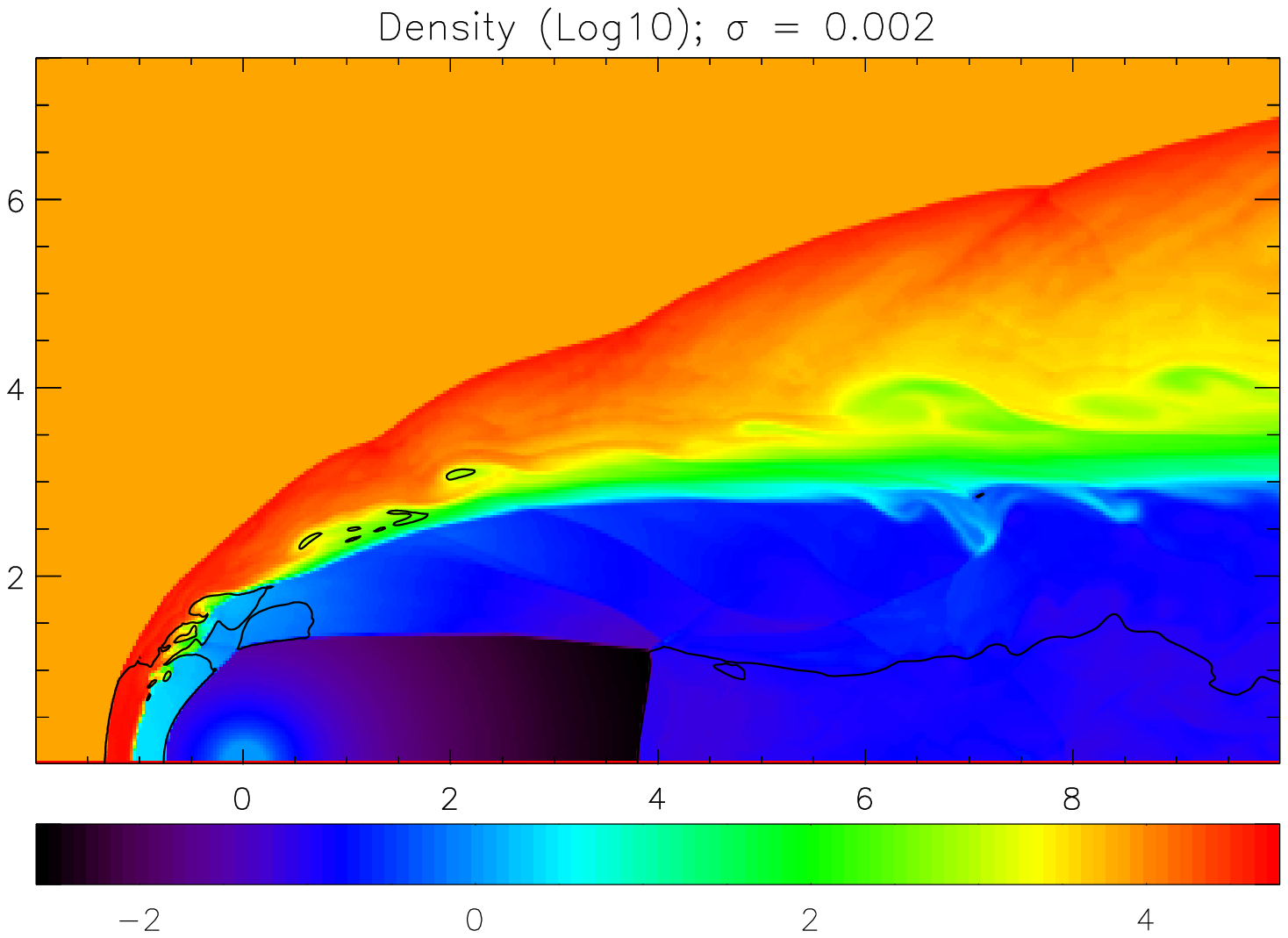}}
\resizebox{\hsize}{!}{\includegraphics{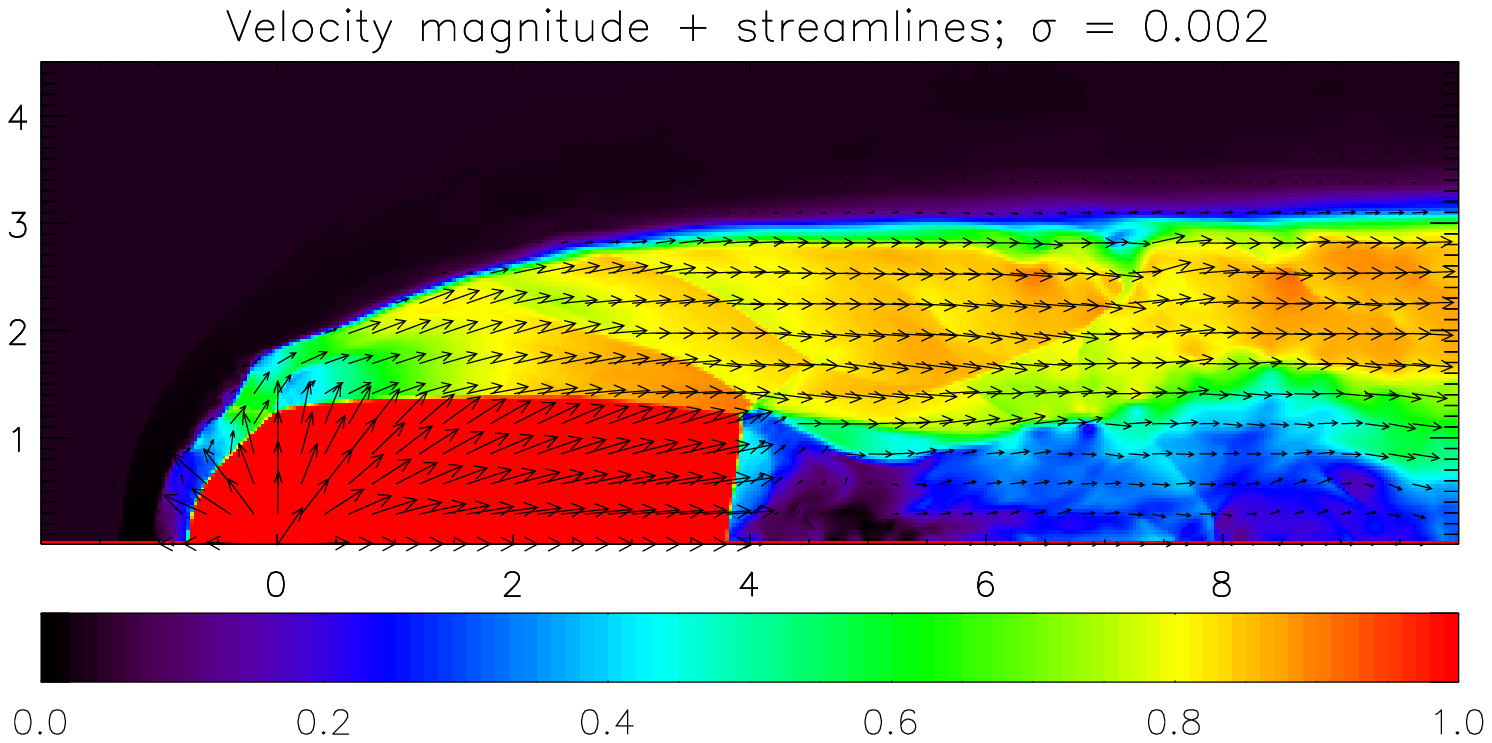}}
\resizebox{\hsize}{!}{\includegraphics{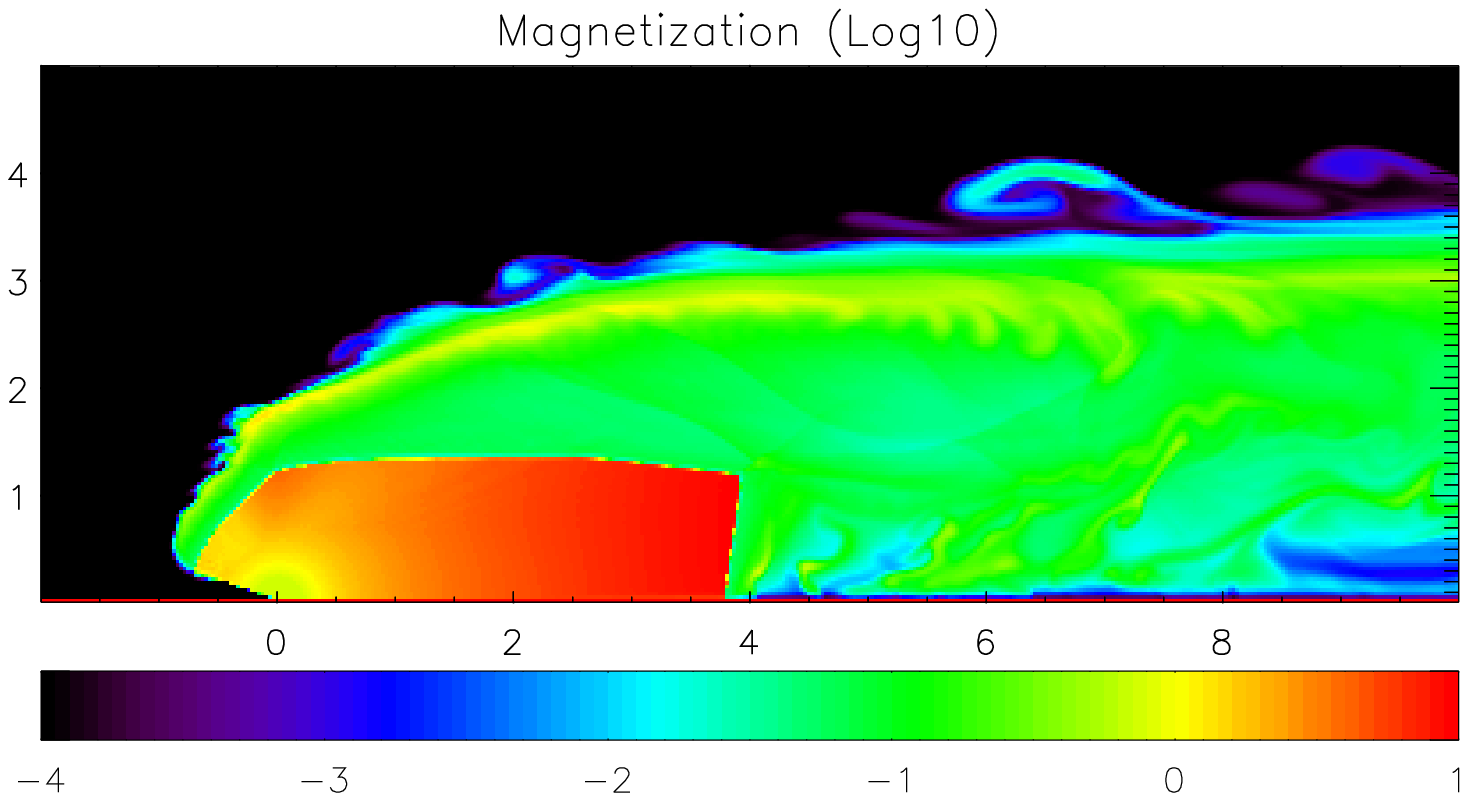}}
\caption{Flow structure in the case $\sigma=0.002$. Upper panel: the color map represents the density in logarithmic scale (arbitrary units); the contours define the sonic surface. Middle panel: the color scale refers to the velocity magnitude; the arrows trace the streamlines. Bottom panel: map of the magnetization in the nebula defined as $B^2/(8 \pi \gamma^2 P)$. Spatial scale is in units of 1.3 $d_o$, corresponding to the distance of the CD from the pulsar, along the axis, in the HD case.}
\label{fig:st1}
\end{figure}

\begin{figure}[h!!!!!!!!!!!!!]
\resizebox{\hsize}{!}{\includegraphics{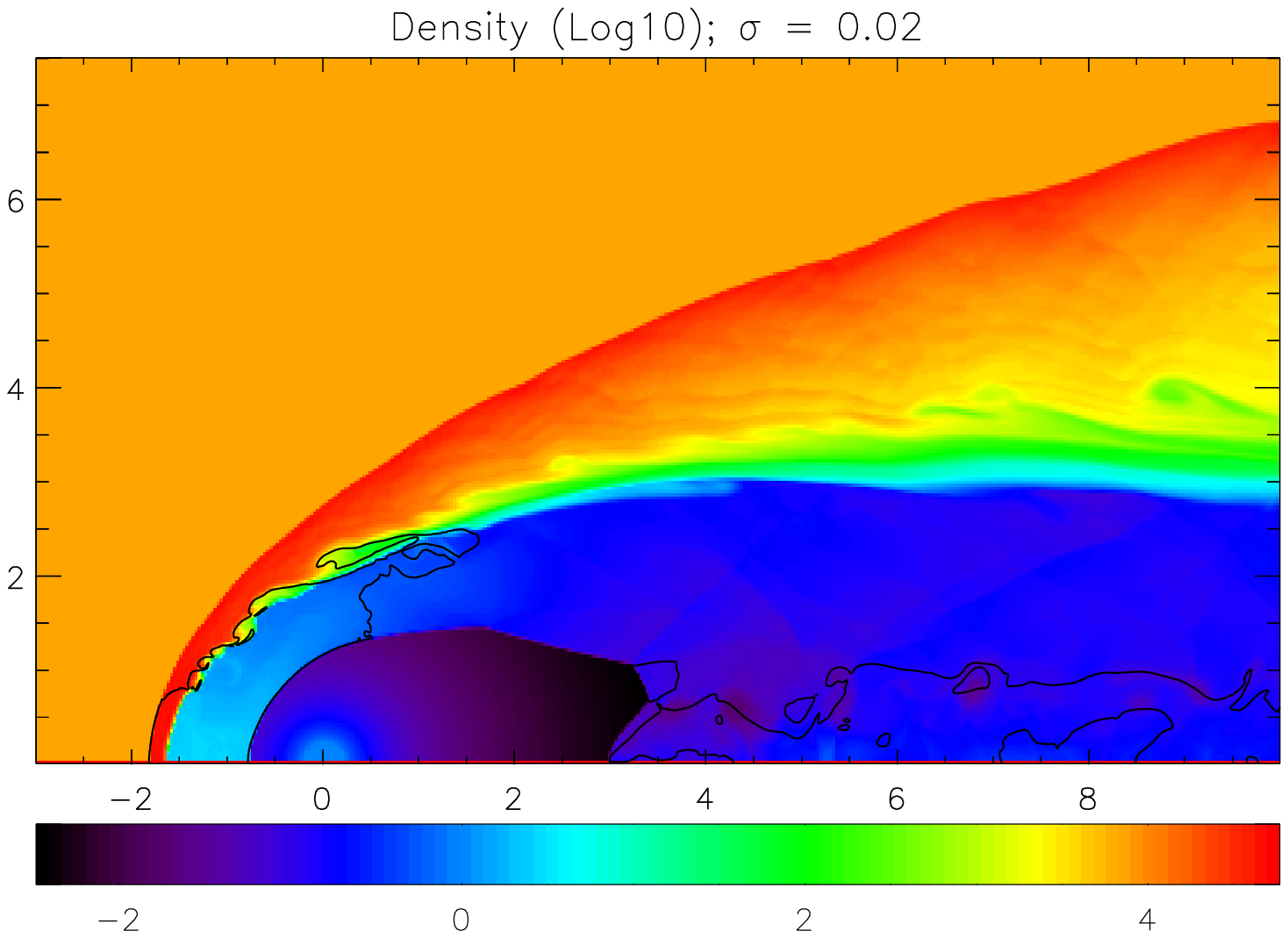}}
\resizebox{\hsize}{!}{\includegraphics{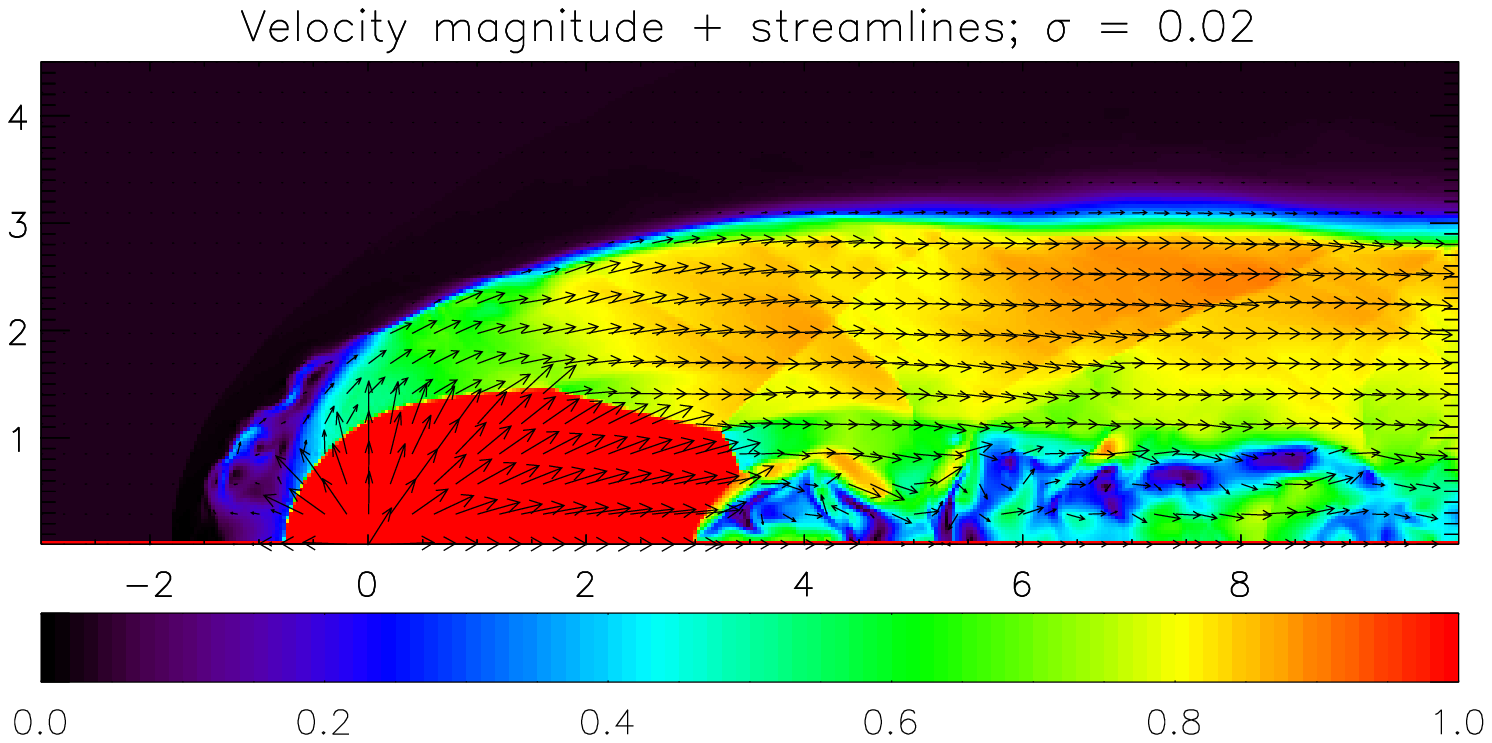}}
\resizebox{\hsize}{!}{\includegraphics{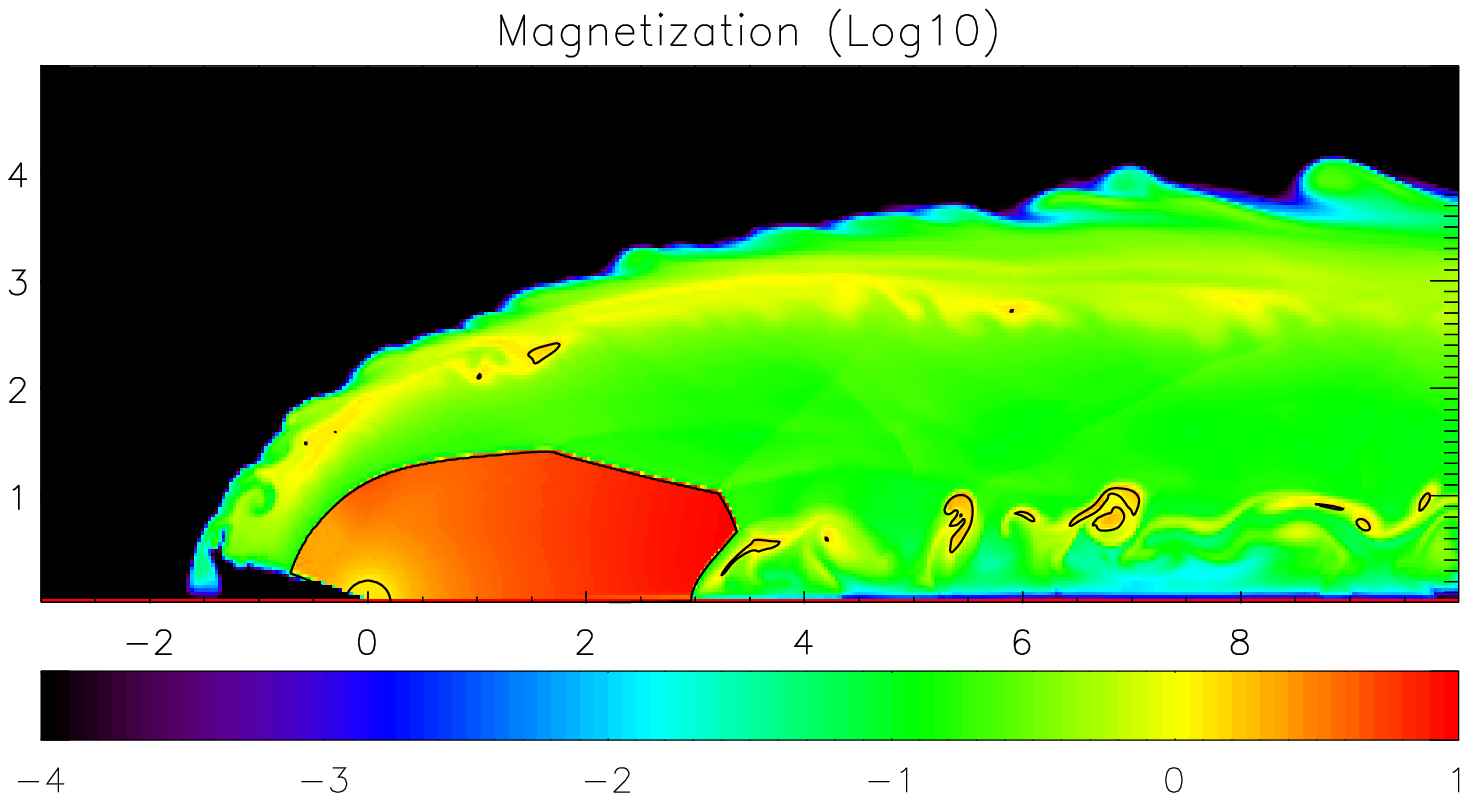}}
\caption{Flow structure in the case $\sigma=0.02$. The legend is the same as for Fig.~\ref{fig:st1}. Now in the bottom panel contours appear, that represent equipartition surfaces.}
\label{fig:st2}
\end{figure}

\begin{figure}
\resizebox{\hsize}{!}{\includegraphics{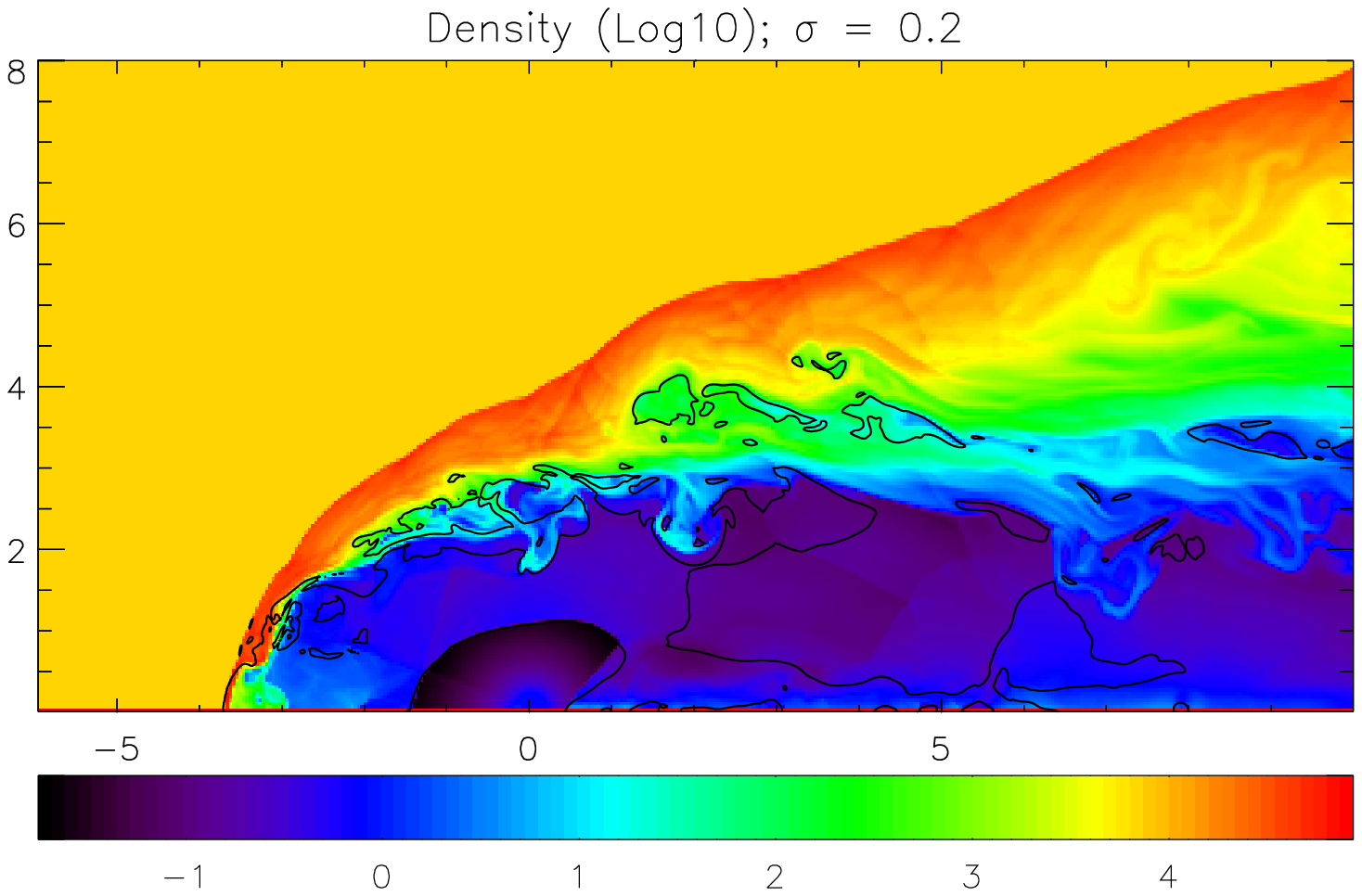}}
\resizebox{\hsize}{!}{\includegraphics{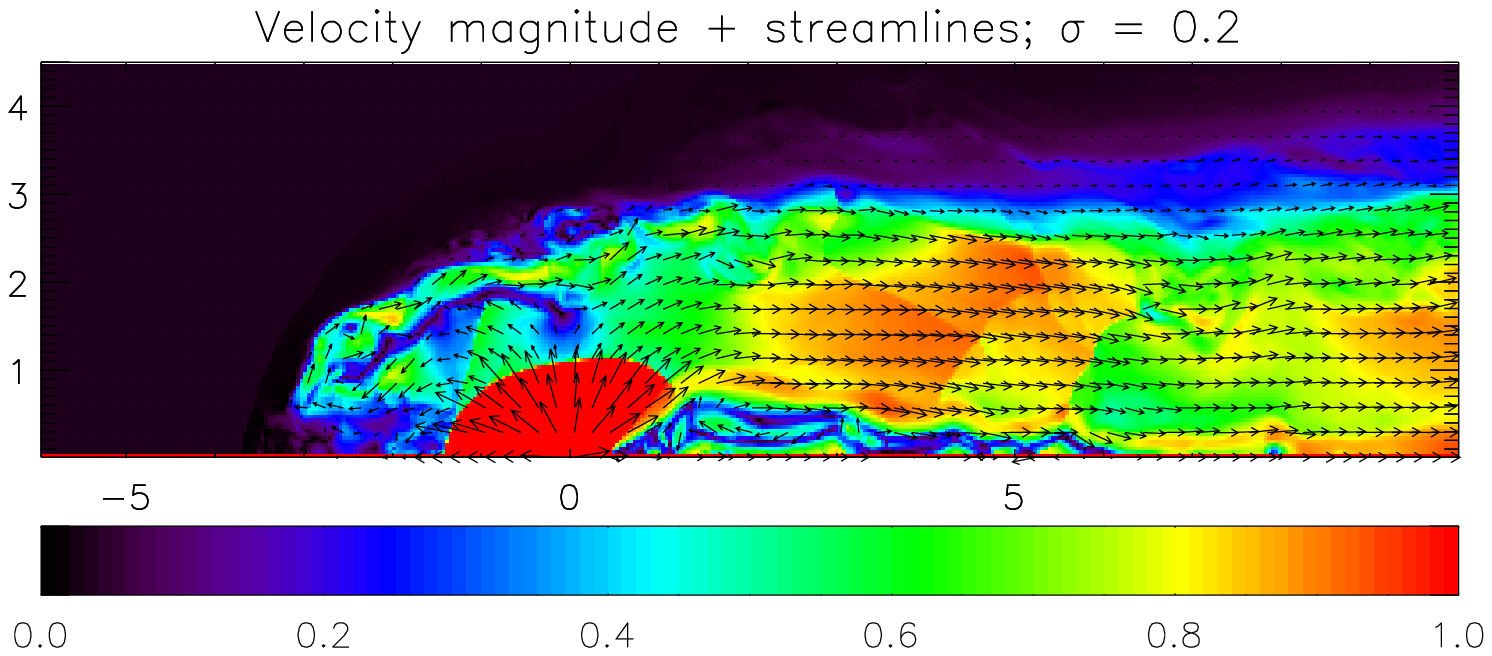}}
\resizebox{\hsize}{!}{\includegraphics{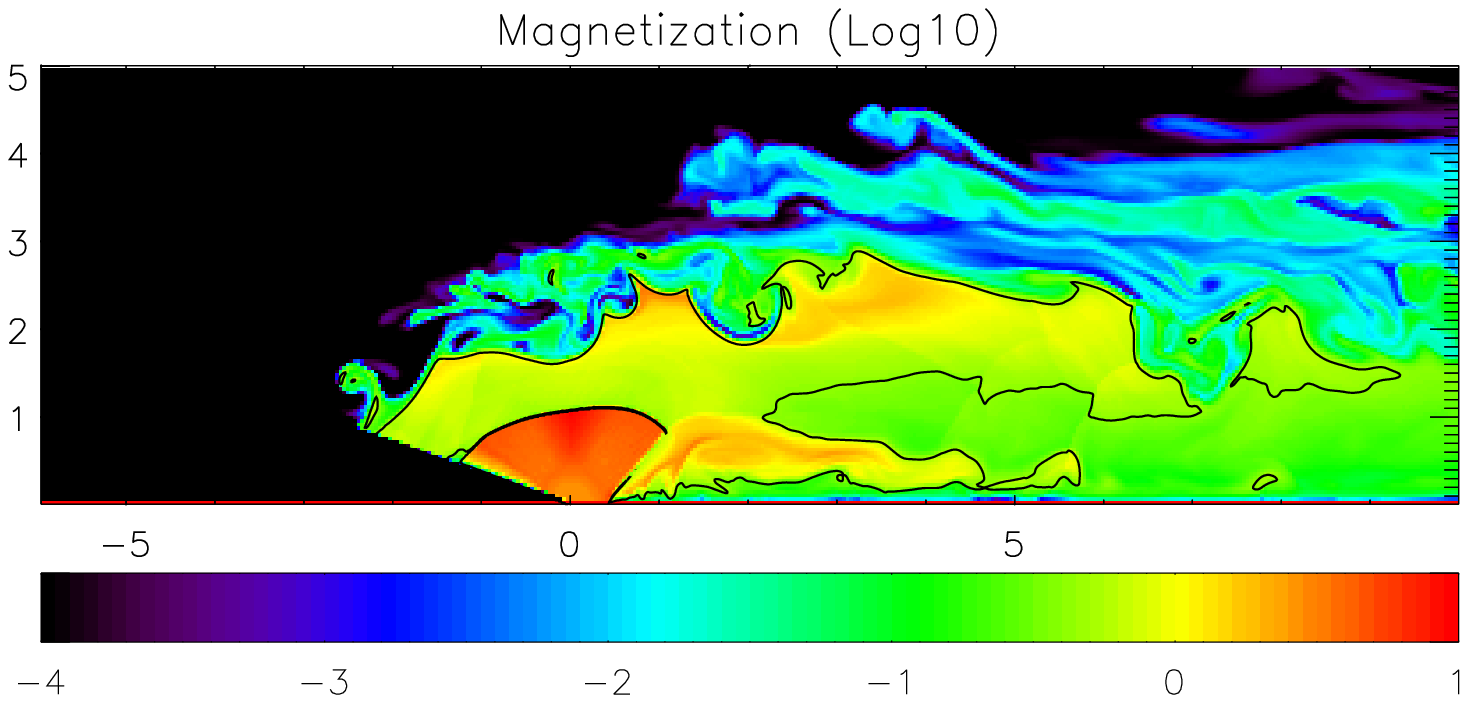}}
\caption{Flow structure in the case $\sigma=0.2$. The legend is the same as for Figs.~\ref{fig:st1}-\ref{fig:st2}.}
\label{fig:st3}
\end{figure}

Regarding the general shape of the inner region occupied by the hot relativistic plasma, these new simulations confirm previous results (Bucciantini \cite{bucciantini02}) suggesting that the material is collimated in a cylindrical tail. The radial extent of this cylinder is found to be $R_{\rm tail} \simeq 4 d_o$ and it does not depend neither on the wind magnetization nor on the Mach number (in the hypersonic limit). This can be easily understood in terms of energy conservation and general scalings for the fluid quantities. In the weakly magnetized cases one can write the energy conservation as:
\be
\dot{E} = 2 \pi \int_0^{R_{\rm tail}} 4\,P_{\rm tail}\,\gamma^2\,v_{\rm tail}\,R\,dR\ ,
\label{eq:econs}
\ee 
with all quantities in the integral depending in general on the cylindrical radius $R$, $v_{\rm tail}$ being the flow speed. Neglecting the radial profiles and using the definition of the stagnation point (Eq.~\ref{eq:stag}):
\be
\frac{P_{\rm tail}}{\rho_o V^2}=\frac{1}{\gamma_{\rm tail}^2} \frac{c}{v_{\rm tail}} \left( \frac{d_o}{R_{\rm tail}} \right)^2
\label{eq:ptail}\ . 
\ee  
The only scale velocity for the plasma inside $R_{\rm tail}$ is the speed of the wind  which is $\sim c$. The ``two thin layers'' solution (Bucciantini \cite{bucciantini02b}) would predict $v _{\rm tail}\sim 0.6 c$. In fact, in our simulations (middle panel of Figs.~\ref{fig:st1}-\ref{fig:st3}) we have, in all cases, $v_{\rm tail}\sim 0.8-0.9 c$ ($\gamma_{\rm tail} \sim 2$). In addition we find, as mentioned above, $P_{\rm tail}\sim 0.02 \rho_o V_o^2$. Substituting these numerical values in Eq.~\ref{eq:ptail} one easily obtains $R_{\rm tail}\sim 4 d_o$.

Considering in greater detail the internal structure of the relativistic plasma flow in the tail, we find a qualitative agreement with previous analyses (Bucciantini \cite{bucciantini02}). The flow has different average properties in two main regions (see middle panel of Figs.~\ref{fig:st1}-\ref{fig:st3}). The inner channel contains particles coming from the BTS. The latter, in the HD and $\sigma=0.002$ cases, is almost perpendicular to the wind propagation direction, so that the flow downstream is subsonic (flow speed $\sim c/3$ in the HD case). The outer channel, surrounding the former, contains particles coming from the head of the nebula and from a portion of the TS which is oblique with respect to the wind speed. The fluid in the head is subsonic due to the shock geometry, but, as it expands sideways, it may undergo a sonic transition and become supersonic, or superfastmagnetosonic in the MHD case (sonic surfaces are represented as contours in the upper panel of Figs.~\ref{fig:st1}-\ref{fig:st3}). The angular distance from the apex at which this happens depends on the magnetization. For low $\sigma$ it takes place at a polar angle $\sim 60^\circ -70^\circ $. As the magnetization increases the transition occurs further away: for $\sigma=0.02$ the angle is $\sim 120^\circ$, while for $\sigma=0.2$ the downstream flow is everywhere subfastmagnetosonic and transitions take place in the tail. 

Typical speeds of the two channels are different. In the supersonic outer channel the velocity is not much dependent on magnetization and is of order $0.8 -0.9 c$. In the subsonic inner channel values of the speed are in the range $0.1-0.3 c$ for $\sigma=0.002$ and in the range $0.2-0.5c$ for $\sigma=0.2$.  The shear between the flow in these two channels is at the origin of Kelvin-Helmoltz type instabilities at their CD. The effects of this instability are not evident in the top panels of Figs.~\ref{fig:st1}-\ref{fig:st3} due to the low density contrast, but with a greater contrast the presence of mixing and dragging from the fast component to the slow one becomes clear. Moving away from the TS, the speed in the slow channel tends to increase as a consequence of this dragging and it may become supersonic.

Shear instability is also present at the CD between the outer layer of shocked ISM and the inner layer of shocked pulsar wind. As we mentioned, the flow speed in the latter is close to $c$, while in the former the fluid velocity is of the same order as the pulsar speed through the ISM, $V$. The large velocity difference makes the development of Kelvin-Helmoltz instability at the interface an obvious expectation. However no signs of it had been observed in previous numerical studies (Bucciantini \cite{bucciantini02}, van der Swaluw \cite{swaluw03}), where the flow in the tail was completely laminar. This may partly be due to the large artificial and numerical diffusion present in those works, and partly be intrinsic to the non-relativistic treatment of the problem. In fact, in the framework of classical HD, the velocity difference at the CD is smaller and the density jump between the two layers is strongly reduced. 

%----------------------------------------- Fig 1
%-----------------------------------------

%----------------------------------------- Fig 2
%-----------------------------------------

In a proper relativistic treatment the ratio between the enthalpy of the outer ($\sim  \rho_o c^2$) and inner layer ($\sim 4P\simeq 4\rho_oV^2$) is $\sim v^2/c^2$. For typical values of pulsar proper velocities $V \sim 300$ \kms this ratio is $10^6$. This implies that even a very weak contamination (1 part in a million) of the inner region by the shocked ISM might substantially affect the internal flow dynamics, while larger contamination can in principle modify the overall nebular shape (Lyutikov \cite{lyutikov03}). In our case fingers and eddies of heavier material (that in the simulation is however about 1000 times less dense than realistic ISM) can detach from the outer layer forming clumps of non relativistic material embedded in the hot magnetized pulsar wind plasma. Simulations show that the instability forms at intermediate latitudes in the portion of the head of the nebula that is still subsonic. In the very head, close to the apex, the shear velocity is smaller and the instability is not present, while in the supersonic part of the tail clumps of material are advected far from the pulsar at a speed $\sim 0.5 c$. These clumps can in principle affect the flow structure in the tail while they are advected away: this effect could show as a modulation in the synchrotron emission maps. 

Shear instability is also at the base of two other phenomena observed in our simulations. First of all, it is responsible for the fluctuations in the shape of the TS. Secondly, it is at the origin of a series of oblique shocks that propagate in the outer supersonic channel. Such oblique shocks are a common feature in supersonic collimated flows, and are found in many jet simulations (see for example Komissarov \cite{komissarov99}). Even if they are weak or only moderately strong, they are important as possible sites of particle acceleration. 

However, we must point out that the setting of our simulations does not guarantee a proper treatment of either of the shear unstable surfaces we see, both between fast and slow channel and between the outer and inner layer. The main limitation arises from the 2D axisymmetric approximation and the absence of a poloidal component of magnetic field that could stabilize the interfaces. We also observe that the instability increases with magnetization, this might in principle be due to the fact that for high $\sigma$ the flow is subsonic in a more extended region, although, once again, numerical problems cannot be completely excluded as a cause. 

\subsection{Magnetic field distribution}

Finally, let us consider the structure of the magnetic field. As it is known from solar wind simulations, the magnetic field at the CD with the outer layer  is compressed. An interesting question is how thick this magnetic sheet is. If the magnetization in the wind is not too high ($\sigma <0.1$) the flow structure is essentially the same as in a HD case. Magnetic field is passively advected and compressed toward the CD where it eventually reaches equipartition. As one can see from the bottom panel of Fig.~\ref{fig:st1} equipartition is reached in a narrow sheet close to the CD. In the lowly magnetized cases the nebular magnetic field is on average well below equipartition. This behavior changes when $\sigma$ is larger than $\sim 0.1$. This corresponds to the value at which the magnetic field in the tail can be estimated to be at equipartition. From flux conservation the magnetic field in the tail of the nebula, $B_{tail}$, is, in the pulsar frame:
\be
B_{\rm tail}=\pi \frac{rB_w(r)}{R_{\rm tail}} \frac{c}{v_{\rm tail}}
\ee
with $B_w(r)$ the magnetic field in the wind at a distance $r$ from the pulsar ($B_w(r)\propto r^{-1}$), and $R_{tail}$ the radial extent of the tail, $R_{\rm tail}\simeq 4d_o$.  When, in the tail, the thermal pressure, $P_{\rm tail}$, is dominant, its average value can be derived from energy conservation, Eq.~\ref{eq:ptail}. Neglecting once again the radial profiles:
\be
P_{\rm tail}=\frac{\dot{E}}{4\pi R_{\rm tail}^2\,\gamma^2\,v_{\rm tail}}.
\ee 
We then obtain for the magnetization in the tail:
\be
\frac{B_{\rm tail}^2}{8\pi P_{\rm tail}\gamma^2}
%=\frac{\pi^2}{8\pi}\frac{r^2B_w^2(r)}{R_{\rm tail}^2}\left(\frac{c}{v_{\rm tail}}\right)^2\frac{4\pi R_{\rm tail}^2 v_{\rm tail}}{\dot{E}}\nonumber\\
=\frac{\pi^2}{2}\frac{c}{v_{\rm tail}}\sigma\simeq 6\,\sigma.
\label{eq:btail}
\ee
When $\sigma \simgt 0.1$, the dynamics in the tail is dominated by the magnetic field. Pressure in the internal region is no longer nearly uniform in the transverse direction, and becomes larger toward the axis, due to magnetic pinching, causing the BTS to recede and become convex in shape. Moreover, while for $\sigma=0.02,0.002$, equipartition is reached only in the vicinity of the CD, now the magnetic field is above the equipartition value also close to the axis. Since synchrotron radiation depends on $B^2$, this change in the magnetic energy profile might give important signatures in the emission maps.

%----------------------------------------- Fig 3
%-----------------------------------------

%%%%%%%%%%%%%%%%%%%%%%%%%%%%%%%%%%%%%%%%%%%%%%%%%%%%%%%%%%%%%% 4
\section{Synchrotron emissivity}

Given the flow structure, magnetic energy distribution and pressure profile inside the nebula, it is possible to compute synchrotron emission maps. Following Begelman \& Li (1992), we assume that particle acceleration takes place only at the TS and the resulting particle energy distribution is:
\be
f(\epsilon_o,\Psi)=\xi\,P_o(\Psi)\,\epsilon_o^{-(2\alpha+1)},
\label{eq:distr}
\ee
with
\begin{eqnarray}
\xi=&3 K_p \left[\int_{\epsilon_m}^{\epsilon_M}\epsilon_o^{-2\alpha} d\epsilon_o\right]^{-1} & \nonumber \\
= &3\,K_p\,(2\alpha-1)\frac{\epsilon_m^{2\alpha-1}}{1-(\epsilon_m/\epsilon_M)^{2\alpha-1}} & {\rm for}\; \; \alpha\neq0.5\ .
\end{eqnarray}
Here $K_p\,P_o(\Psi)$ is the fraction of the thermal pressure downstream of the TS that is carried by accelerated particles, $\epsilon_m$ and $\epsilon_M$ are the minimum and maximum energy of the particle spectrum respectively, and $\Psi$ labels the different streamlines diverging from the TS. $P_o$ is different all along the TS but since the shock is everywhere strong, we assume$K_p$, $\epsilon_m$ and $\epsilon_M$ to be independent on the streamline. Moreover in the following we will always assume $\epsilon_m \ll \epsilon_M$ and $\alpha \neq 0.5$.

At a generic distance $s$ along a streamline the particle distribution function, evolved taking into account adiabatic and synchrotron losses, is :
\begin{eqnarray}
f(\epsilon,s,\Psi)&=&\xi\; P(s,\Psi)\; \epsilon^{-(2 \alpha+1)} \nonumber \\
&\times&\left(P(s,\Psi)/P_o(\Psi)\right)^{\frac{2\alpha-1}{4}} \left(1-\epsilon/\epsilon_\infty\right)^{2\alpha-1},
\label{eq:evdistr}
\end{eqnarray}
with
\be
\epsilon_\infty=\epsilon_\infty(s,\Psi)=\rho^{1/3}\left[\frac{4}{9}\frac{e^4}{m^4c^7}\int_o^s\frac{B^2 \rho^{1/3}}{v}ds\right]^{-1}\,
\ee
where $\rho$ is the relativistic plasma density.
As we will discuss below, for the typical range of pulsar velocities and luminosities, synchrotron losses are negligible within the spatial region corresponding to our computational domain and for the frequency range we are interested in. This fact translates into $\epsilon_\infty\gg \epsilon$, that allows one to simplify  Eq.~\ref{eq:evdistr}.

Another effect that needs to be taken into account is Doppler boosting, since the emitting plasma is moving with bulk velocity close to $c$. We decided for simplicity to consider a nebula being viewed face on, i.e. with the pulsar speed laying in the plane of the sky. In calculating the volume emissivity $s(\nu)$ toward the observer we have approximated the radiation spectrum of the single particles as a Dirac delta centered on the critical frequency $\nu_c=c_2 B \epsilon^2$. In this approximation we obtain:
\be
s(\nu)=\frac{c_1\; c_2^{\alpha-1}}{2}\; \xi\; P\; D^{2+\alpha}\left(P/P_o\right)^{\frac{2\alpha-1}{4}}(B'_\bot)^{\alpha+1}\;\nu^{-\alpha}
\label{eq:snu}
\ee
where $c_1=(4/9)e^4c/(mc^2)^4$, $c_2=0.29 \times 3ec/(4\pi(mc^2)^3)$, $B'_\bot$ is the magnetic field component perpendicular to the observer's line of sight, as measured in the fluid frame, and $D=(\gamma(1-{\bf n}\cdot \Bbeta))^{-1}$ is the Doppler factor. Here $\Bbeta$ is the flow velocity and ${\bf n}$ is the unit vector pointing toward the observer, both in the pulsar frame. The latter is related to the observer direction in the fluid frame ${\bf n}'$ through (see e.g. Komissarov \& Lyubarsky \cite{komissarov04}):
\be
{\bf n}'=D\left({\bf n}+\Bbeta\left[\frac{\Bbeta\cdot {\bf n}}{\beta^2}(\gamma -1) -\gamma\right]\right).
\ee

Before computing the emission map at a given frequency we need to choose a spectral index $\alpha$. Estimates of the X-ray spectral index are presently available only for four pulsar bow shock nebulae: for IC443, Bocchino \& Bykov (\cite{bocchino03}) find $\alpha \le 0.5$; for the Duck nebula Kaspi et al. (\cite{kaspi01}) give $\alpha=0. \pm 0.6$; for PSR B1957+20 Stappers et al. (\cite{stappers03}) find $\alpha=0.9 \pm 0.5$; finally, Gansler et al. (\cite{gaensler04}) give for the Mouse nebula $(0.8 \pm 0.1) \le \alpha \le (1.6 \pm 0.2)$. For our simulated map we decided to use $\alpha=0.6$, a choice which avoids the critical value of $\alpha=0.5$ and leaves in Eq.~\ref{eq:snu} only a very weak dependence on the conditions at the TS ($P_o$).  

\begin{figure}
\resizebox{\hsize}{!}{\includegraphics{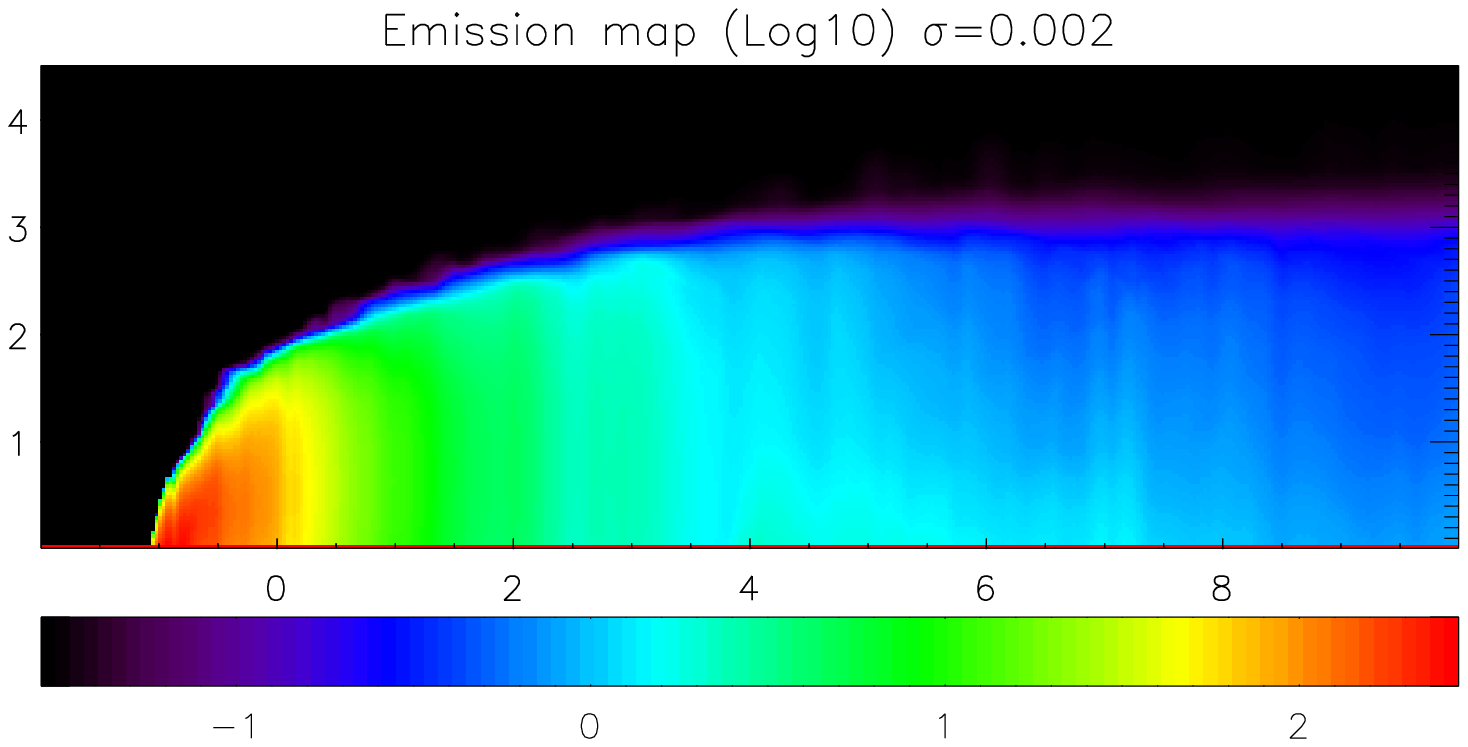}}
\resizebox{\hsize}{!}{\includegraphics{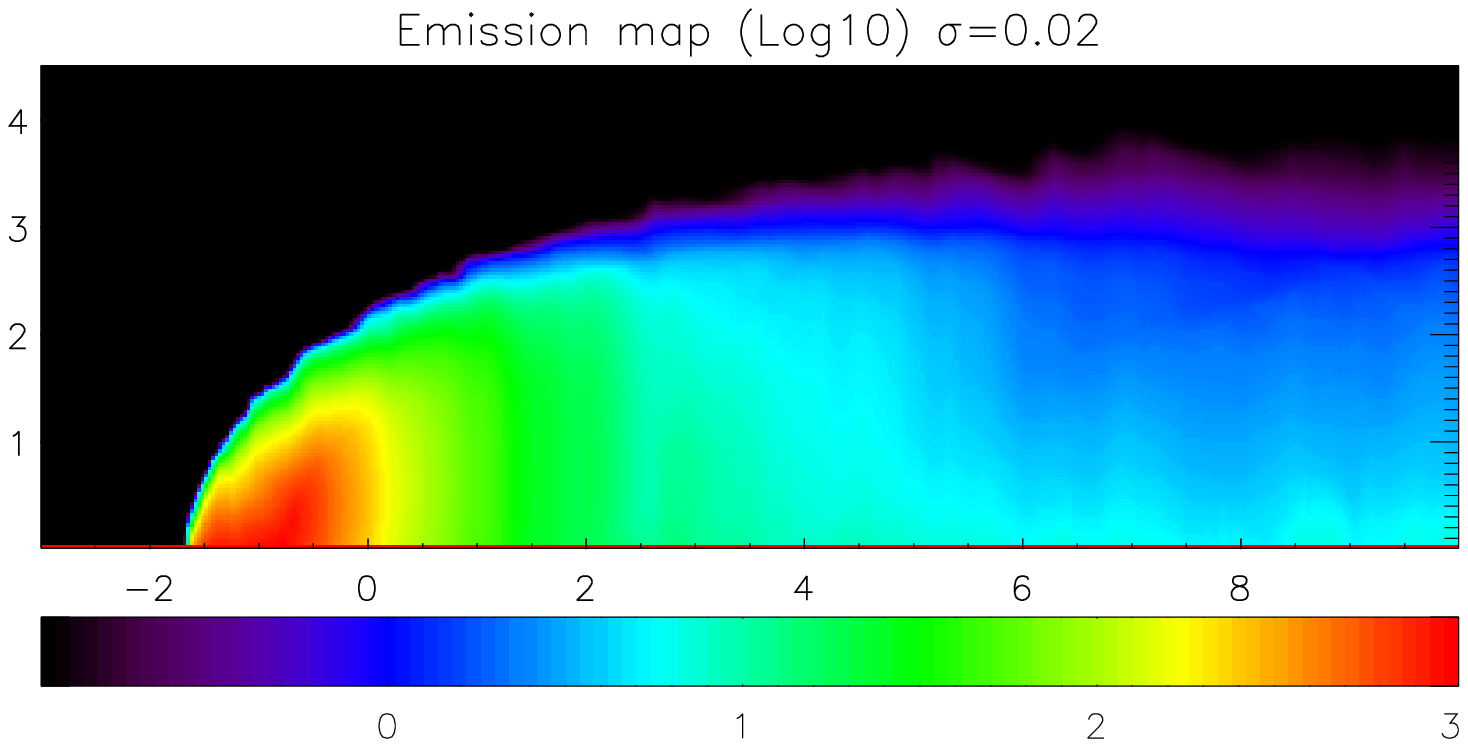}}
\resizebox{\hsize}{!}{\includegraphics{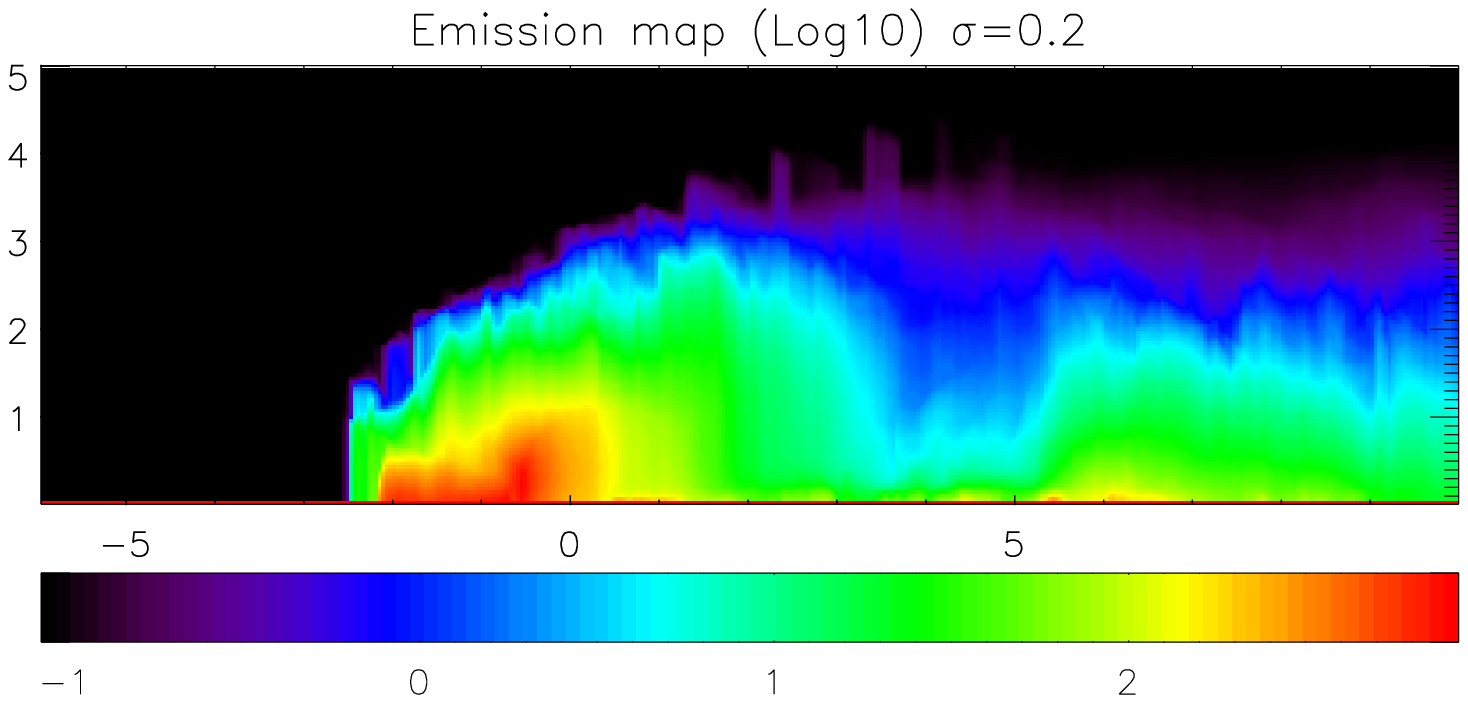}}
\caption{Simulated synchrotron maps corresponding to the three simulations presented in Sect.~3. The color levels represent the intensity in logarithmic scale (arbitrary units). The spatial scale is the same as in Figs.~\ref{fig:st1}-\ref{fig:st3}. No disordered component of the magnetic field is included in computing the emission. From top to bottom: $\sigma=0.002,0.02,0.2$.}
\label{fig:em1}
\end{figure}

The synchrotron emission maps, that we built by integrating the volume emissivity in Eq.~\ref{eq:snu} along the line of sight, are shown in Fig.~\ref{fig:em1} for all the different values of $\sigma$ we adopted. The brightest part of the nebula corresponds to the head of the bow-shock, where magnetic field and particle pressure are highest. Here Doppler boosting is not very important since the flow velocities involved are low. Moving from the head toward the back of the nebula we find a region of fainter emission and then an even fainter tail. It is interesting to notice that in the case of the Mouse nebula Gaensler et al. (\cite{gaensler04}) distinguish three main regions in the X-ray emission pattern, that they name as the ``head'', the ``tongue'' and the ``tail'', with intensity ratio $100/10/1$. This is the same intensity ratio we find in our synthetic maps between the corresponding regions. The ratio between the head and the tongue, moreover, turns out to be the same for all values of $\sigma$. This would imply that the head/tongue emission properties do not provide us with information on the wind magnetization.

\begin{figure}
\resizebox{\hsize}{!}{\includegraphics{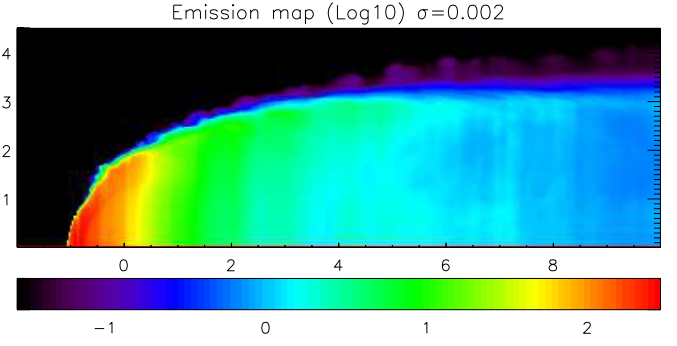}}
\caption{Same as upper panel of Fig.~\ref{fig:em1} but now the amount of magnetic energy that is found at each place in the simulation is assumed to be in the form of a completely disordered magnetic field.}
\label{fig:em2}
\end{figure}

 We also notice that the tail of emission in the back of the nebula is more enhanced toward the axis than close to the CD. This is a consequence of the toroidal structure of the field. Close to the axis the magnetic field lines are perpendicular to the observer and $(B'_\bot)$ is maximum, while at the edge of the inner region the magnetic field is almost pointing toward the observer, so that  $(B'_\bot)$ reaches a minimum. However, as we stressed in the previous section, shear instabilities both at the CD and between the fast and slow internal channels can in principle destroy the ordered structure of the magnetic field. If a disordered component of the magnetic field is introduced in Eq.~\ref{eq:snu}, the contrast between the axis and the edges weakens. In Fig.~\ref{fig:em2} we show the synchrotron map corresponding to the case $\sigma=0.002$ when a completely disordered magnetic field is used in Eq.~\ref{eq:snu}, while keeping the magnetic energy distribution inside the nebula the same. 

 The difference in the axis-edge intensity contrast between an ordered and a disordered magnetic field becomes smaller with increasing $\sigma$. This is because, as magnetic stresses in the tail become dynamically more important, the magnetic energy density becomes higher close to the axis, where projection effects are less important. For the same reason, in comparison with cases at lower $\sigma$, for $\sigma=0.2$ the emission is enhanced toward the axis (bottom panel of Fig.~\ref{fig:em1}). We suggest that the observation of an emission tail narrower than the head of the nebula might be considered a signature of high ($>0.1$) $\sigma$ winds. However a high $\sigma$ wind is not bounded to produce such a structure, because this is strictly related to the axisymmetric approximation. In fact, if the spin axis is not aligned with the pulsar velocity, the magnetic energy distribution is going to be different. Moreover, even in the case of alignment, it is always a possibility that 3D instabilities disrupt the ordered magnetic field structure eliminating hoop stresses.

As we discussed in the previous section, the shear at the CD is at the origin of instabilities and causes the formation of dense clumps of unmagnetized shocked ISM, that are embedded in the relativistically hot plasma. Such instabilities, that appear to be stronger at high magnetization, and the clumps they generate, can in principle modulate the emission in the tail, both distorting the fluid flow pattern and compressing the relativistic material. Typical time scales for the detachment of clumps and their flow time in the tail are of order $d_o/c$. Evidence of modulation in the emission pattern of non thermal radiation in pulsar bow-shock nebulae has been found in the case of the Mouse. Unfortunately we have no information about the magnetic field structure in this nebula (i.e. pulsar spin axis inclination) and/or  time variability of the observed modulation that could confirm or rule out theories on its origin. In fact other possible explanations have been suggested so far that do not require instability at the CD, as expansion-compression waves (Lyutikov \cite{lyutikov04}).

As a final remark on the spatial structure of the emission let us shortly discuss the dependence of the results on our assumption for the particle spectrum and for the orientation of the nebula. In the case of a particle distribution function with $\alpha$ much different from 0.5, in building emission map the conditions at the TS would play a much more relevant role (see Eq.~\ref{eq:snu}). In general the effect is to enhance the contrast between the vicinities of the pulsar and the edges of the nebula. However, even in the case of particles coming from the head, where the pressure is highest, and flowing in the tail, the correction is just a factor $\sim 2$ for $\alpha=1$. Much more dramatic changes in the emission pattern may come instead from a different orientation of the pulsar velocity. If the pulsar proper motion has a non-negligible component out of the plane of the sky, relative motions in the tail become important. For a pulsar moving toward the observer the emission in the tail is weaker than what is shown in Fig.~\ref{fig:em1}, while in the case of a pulsar moving away from the observer emission should be enhanced.

Let us focus now on the issue of synchrotron losses. In the above discussion we have assumed that particles of all energies, within the range of interest for emission in the X-rays, survive everywhere in the computational domain. An estimate of the synchrotron lifetime might be easily given. The magnetic pressure in the the tail of the nebula is approximately given by Eq.~\ref{eq:btail}, and, using Eq.~\ref{eq:ptail}, can be written as:
\begin{eqnarray}
B^2&&\simeq 4\pi^3\left(\frac{c}{v_{\rm tail}}\right)^2\left(\frac{d_o}{R_{\rm tail}}\right)^2\sigma\; \rho_o V^2\nonumber\\
&&\simeq7.7\left(\frac{c}{v_{\rm tail}}\right)^2\sigma\; \rho_o V^2\ .
\label{eq:bbtail}
\end{eqnarray}
We approximate the age of a particle found at a distance $\eta_d d_o$ from the apex as $\sim (\eta_dd_o)/(\eta_vc)$ where $\eta_v c$ is the average flow speed the particle has experienced. In the following we will be interested only in particles flowing in the tail and with and average velocity such that $\eta_v \simeq v_{\rm tail}/c$: this implies that we are not considering particles accelerated at the FTS, within a polar angle of about $\pi/4$ from the axis. Given the magnetic field in Eq.~\ref{eq:bbtail}, the maximum energy of the photons emitted by this particle turns out to be:
\begin{eqnarray}
\epsilon_b\simeq&&3\times10^{8}\ \sigma^{-\frac{3}{2}}\, \eta_d^{-2} \left(\frac{v_{\rm tail}}{c}\right)^5\left(\frac{\rho_o}{10^{-24}{\rm g\, cm^{-3}}}\right)^{-\frac{1}{2}}\nonumber\\
&&\left(\frac{V}{10^7{\rm cm\,s^{-1}}}\right)^{-1}\left(\frac{\dot{E}}{ 10^{34}{\rm erg/,s^{-1}}}\right)^{-1} {\rm keV}.
\label{eq:eb}
\end{eqnarray}

As already pointed out, the value $v_{\rm tail}\simeq 0.8-0.9 c$ is independent of the magnetization. Therefore, knowing the extent of the tail of the nebula at a given frequency, the above relation (Eq.~\ref{eq:eb}) allows one to estimate the wind magnetization $\sigma$.

\subsection{The mouse}
As a special case let us consider the bow-shock nebula around PSR J1747-2958 (Gaensler et al.~\cite{gaensler04}), also called the Mouse Nebula. As we mentioned in the introduction, this is the only bow-shock nebula for which data are available both at radio and X-ray frequencies. We now wish to discuss these data in the light of our numerical results. 

It has been suggested that, in this nebula, synchrotron losses are responsible for the different morphologies observed in these two bands.

First of all, let us consider the most recent X-ray results concerning the extent of the nebula. Gaensler et al (\cite{gaensler04}) give for the length of the tail $l_{tail}\simeq 1.1\ {\rm pc}$, and for the stagnation point distance $d_o \simeq 0.024\ {\rm pc}$, for an assumed distance of 5 kpc. Using Eq.~\ref{eq:eb} with $\epsilon_b=2.5\ {\rm keV}$, $\dot{E}=2.5 \times 10^{36} {\rm erg\,s^{-1}}$, $V_o=6 \times 10^7 {\rm cm\,s^{-1}}$ and $\rho_o = 0.3 \times 10^{-24} {\rm g\,cm^{-3}}$, we find:
\be
\sigma \simeq 20 \left(\frac{v_{\rm tail}}{c}\right)^{10/3}\ .
\label{eq:sigv}
\ee  
As we mentioned above, our simulations give for the flow speed in the tail $v_{\rm tail}/c \simeq 0.8-0.9$. This implies that the above relation cannot be satisfied for $\sigma < 1$. It must be stressed, though, that Eq.~\ref{eq:sigv} only holds in the low $\sigma$ limit, {\it i. e.} $\sigma\simeq 0.1$ (see Eq.~\ref{eq:btail}), and, even using the upper limit, we find that the velocity should be $v_{\rm tail} < 0.2 c$, in order to obtain the observed value for the X-ray extent of the nebula.

Such a value of the velocity is much lower than what we find in our ideal MHD models. A means of slowing down the flow efficiently is possibly through contamination by the external non-relativistic ambient medium. This could in principle provide at the same time an explanation for another discrepancy we find between our ideal MHD picture and the data, {\it i.e.} the radial extent of the tail at radio frequencies. The latter in fact is about a factor 3-4 larger than what we find, although preliminary results of an investigation currently underway show that this discrepancy could be alleviated by introduction of a latitude dependence of the wind energy flux (analogous to what is done for Crab-like plerions, e.g. Del Zanna et al., \cite{delzanna04}). The flow divergence can also justify the progressive steepening of the X-ray spectrum moving away from the pulsar along the nebula, as well as the smaller radial extent of the X-ray tail than at radio wavelengths. However a self-consistent analysis of the effects of mass loading is beyond the scope of this article. 

Using Eqs.~\ref{eq:distr}-\ref{eq:snu} it is also possible to evaluate the expected value of $L_x/\dot{E}$ for the different values of $\sigma$ we employed, and to compare our results with the value measured for the Mouse Nebula, $L_x/\dot{E}=0.02$. We built synthetic maps using the same algorithm described above but with $\alpha=1$, according to the photon index measured by Gaensler et al. (\cite{gaensler04}),  and extrapolating the emissivity in the tail to a distance $\sim 45 d_o$ from the pulsar. Integration was performed in the {\it Chandra} band, from 0.5 to 5 keV. This allows one to fix the normalization $K_p\,\epsilon_{min}$ in the distribution function and the results are:
\begin{eqnarray}
K_p\,\epsilon_{min}&&\simeq10^7\,(L_x/\dot{E}){\rm erg}\;\;\;{\rm for}\;\sigma=0.002, \nonumber \\
K_p\,\epsilon_{min}&&\simeq10^6\,(L_x/\dot{E}){\rm erg}\;\;\;{\rm for}\;\sigma=0.02,\\
K_p\,\epsilon_{min}&&\simeq5\times10^5\,(L_x/\dot{E}){\rm erg}\;\;\;{\rm for}\;\sigma=0.2\ . \nonumber 
\end{eqnarray}
To give an estimate of the fraction of the total energy output from the pulsar that must be converted into X-ray emitting particles in order to give the observed $L_x/\dot{E}$, one can assume  $\epsilon_{min}\simeq 2200\,{\rm erg}$, a value corresponding to a particle emitting a photon of 1 keV in a magnetic field $B^2=8\pi\rho_o V^2$. We find that even at high magnetization a substantial fraction ($\sim 30\%$) of the total energy output from the pulsar must be converted into X-ray emitting particles. This also implies that the transition between the flat ($\alpha \sim 0$) and steep ($\alpha \sim 1$) part of the accelerated particle distribution might be close to the X-ray band. Again, slowing down of the flow with respect to our findings and the consequent compression of the magnetic field might make such constraint somewhat weaker.

%%----------------------------------------- Fig 4
%-----------------------------------------

%%----------------------------------------- Fig 4
%-----------------------------------------

%%%%%%%%%%%%%%%%%%%%%%%%%%%%%%%%%%%%%%%%%%%%%%%%%%%%%%%%%%%%% 5
\section{Conclusion}

In this paper we have presented the first effort to investigate the structure of pulsar bow-shock nebulae by means of relativistic MHD simulations. Our analysis focused on the signatures of the pulsar wind magnetization, $\sigma$, on the flow structure and on the non-thermal emission by the system. 

We found that the structure of the external layer of shocked ISM does not change significantly with varying $\sigma$, apart from the very head of the nebula. Here magnetic compression is stronger and the pile-up of magnetic loops tends to deform the rounded HD shape into a pointed one. We can thus conclude that the \halpha emission should not be much affected by the magnetization of the pulsar wind. While what indeed changes with $\sigma$ is the distance of the front bow-shock, which increases with magnetization. This has to be taken into account when using pressure equilibrium to evaluate the ambient medium density. However, we want to point out that all the effects related to magnetic pile-up, especially close to the axis, might be artificially amplified in our simulations by the assumption of axisymmetry, that prevents the development of kink instabilities.

We found, as expected, that the shape of the TS is sensitive to $\sigma$. At low $\sigma$ it tends to be elongated as in the HD case, with the distance of the BTS from the pulsar $\sim 5$ times that of the FTS. With increasing magnetization this ratio decreases as a consequence of the internal pressure profile. 

In the tail, our simulations confirm the HD result that the pressure at the CD saturates at a value $\sim 0.02\rho_oV^2$, and the expectation that the magnetization increases $\propto\sigma$. In the low $\sigma$ cases the magnetic field is always below equipartition and it tends to peak toward the CD. On the contrary, for $\sigma>0.1$, magnetic stresses in the tail are important, the magnetic field peaks close to the axis and exceeds equipartition in most of the nebular volume.

The internal flow in the tail is structured in two main regions: an internal subsonic channel with flow speed $\sim 0.2-0.4 c$ and an outer fast supersonic channel where velocity reaches $0.9c$. The average flow speed is a substantial fraction of the wind speed: we find $v_{\rm  tail}\simeq 0.8-0.9c$, to be compared with the value of $0.6c$ predicted by the ``two thin layers'' solution (Bucciantini \cite{bucciantini02b}).

Emission maps are computed for each flow structure we found in our simulations. The main difference when varying $\sigma$ is that, when the wind magnetization is such that most of the nebula is above equipartition, the emission is enhanced close to the axis. Another point that is worth mentioning is that we see modulations of the emission pattern associated with shear instabilities in the nebula. In general, the emission maps show three main regions, that we denote as a ``head'', a ``tongue'' and a ``tail'', in analogy with the labels used for the Mouse nebula Nebula. Between these regions the intensity scales, in the order, as 100/10/1, a result that is also in qualitative agreement with observations of the Mouse Nebula. 

Moreover, the nebular extent at X-ray energies can be used to constrain the properties of the plasma in the tail, in terms of average speed and magnetization. Comparing our results with what is observed in the case of the Mouse Nebula we conclude that a low $\sigma$ wind requires flow velocities in the tail $v_{\rm tail}\ll c$. It is not clear how such values can be obtained within the framework of an ideal MHD model. On the other hand, even assuming a $\sigma$ that guarantees equipartition in most of the nebula, some non ideal process is still required to lower the flow speed with respect to the values we find. 

Before concluding, let us stress once more the intrinsic limitations of our modeling. First of all, the axisymmetric approximation allowed us to study only configurations in which the speed and spin axis of the pulsar are aligned. In addition, the removal of kink instabilities from the system induces a fictitious pile-up of magnetic field that forced us to include ``magnetic dissipation'' in the head in the high $\sigma$ cases. Finally, we would like to remind that for the correct modeling of these systems, a non ideal process like mass loading is likely to play a key role.

%%%%%%%%%%%%%%%%%%%%%%%%%%%%%%%%%%%%%%%%%%%%%%% ACKNOWLEDGEMENTS
\begin{acknowledgements}
We wish to thank Rino Bandiera for fruitful discussion on the topic. This work has been partly supported by the Italian Ministry for University and Research (MIUR) under grant Cofin2002, and partly by the Italian Space Agency (ASI).  
\end{acknowledgements}

%%%%%%%%%%%%%%%%%%%%%%%%%%%%%%%%%%%%%%%%%%%%%%%%%%%%%%%%%%%% BIB

\end{document}